# Confusion around the tidal force and the centrifugal force

Takuya Matsuda, NPO Einstein, 5-14, Yoshida-Honmachiu, Sakyo-ku, Kyoto, 606-8317
Hiromu Isaka, Shimadzu Corp., 1, Nishinokyo-Kuwabaracho, Nakagyoku, Kyoto, 604-8511
Henri M. J. Boffin, ESO, Alonso de Cordova 3107, Vitacur, Casilla 19001, Santiago de Cile, Chile

We discuss the tidal force, whose notion is sometimes misunderstood in the public domain literature. We discuss the tidal force exerted by a secondary point mass on an extended primary body such as the Earth. The tidal force arises because the gravitational force exerted on the extended body by the secondary mass is not uniform across the primary. In the derivation of the tidal force, the non-uniformity of the gravity is essential, and inertial forces such as the centrifugal force are not needed. Nevertheless, it is often asserted that the tidal force can be explained by the centrifugal force. If we literally take into account the centrifugal force, it would mislead us. We therefore also discuss the proper treatment of the centrifugal force.

## 1. Introduction

In the present paper, we discuss the tidal force. We distinguish the tidal force and the tidal phenomenon for the sake of exactness. The tide is a geophysical phenomenon produced by the tidal force. The surface of the Earth's sea deviates from a spherical shape, leading to the tides. In order to calculate the tidal flow precisely, we have to also take into account Earth's gravity, the centrifugal and Coriolis forces due to the rotation of the Earth, the pressure gradient of the seawater and the friction between the seawater and the sea floor. In addition, we have to take into account the shape of the coastline. The tide is a very complicated phenomenon. The purpose of the present paper is to discuss the tidal potential, and the tidal phenomenon is not discussed.

The notion of the tidal force was discovered by Isaac Newton, and its definition is well understood, with no room for ambiguity in current days. Nevertheless, its explanation depends on authors and some misconceptions creep into the public domain literature. This may be seen, for example, in Wikipedia [1, 2].

The tidal force is the effect of gravity on a body, caused by the presence of a secondary body. On Earth, the tidal force is caused by the Moon and the Sun. The mathematical description of the tidal force is clear and a mathematical discussion is presented in sections 3 to 5.

Since the tidal force was discovered to explain the tides, the system of the Earth and the Moon has been used to explain the tidal force. In such an explanation, the treatment of the centrifugal force may cause some confusion.

In the present paper, we ignore the effect of the Sun. To make the argument simple, the Earth and the Moon are assumed to revolve about a common center of gravity in a circular orbit, the period of which is about one month. Since the mass of the Earth is much larger than that of the Moon, the common center of gravity resides inside the Earth, although this fact is not essential in the present discussion. It is to be noted that the rotation of the Earth about its rotation axis does not affect the tidal force.

The aim of the paper is to derive the tidal force and its potential in the system of the Earth and the Moon. They are produced by the combined effect of the gravity of the Earth and the Moon, and the centrifugal force originated from the rotation of the system of the Earth and the Moon in an inertial frame. Special attention is paid to the understanding of the role of the centrifugal force.

## 2. Confusion around the centrifugal force

As was described earlier, the Earth and the Moon revolve about the common center of gravity in the inertial frame. Fig. 1 shows a wrong picture of the sea surface affected by the tidal force due to the Moon. In the figure, the Earth is shown as the brown circle, the deformed sea surface as the blue ellipsoid and the Moon as the black circle. We consider only the plane on which the Moon revolves in this figure. Along the line connecting the Earth's and the Moon's centers, the points N and F represent those of the near- and the far-side of the Earth's surface, respectively. Note that N and F are fixed to the co-rotating frame and not fixed on the Earth's surface. The Earth rotates in this frame, the period of which is about one day. But this rotation does not affect the tidal force. The green arrows depict the gravity due to the Moon at N and F.

The gravity due to the Moon is larger at N than F as is shown in the Fig. 1. This difference in the strength of the gravity is the cause of the tidal force. At first glance, one may expect that the sea' surface deviates from a circular form to an ellipsoidal one, the center of which is shifted to the Moon from the Earth's center. Of course, this is a wrong picture, given the fact that there are two high tides in a day.

Fig. 2 shows the correct picture of the sea' surface, which is symmetrical about the Earth's center. At the time of Newton, there was a debate about which picture was correct. Nowadays it is obvious that Fig. 2 gives the correct picture, but the problem is 'why is it so?'

This can be explained by Fig. 3, in which the green arrows depict the gravity due to the Moon as before. The black arrows show the centrifugal force due to the rotation of the Earth-Moon system. The magnitude of the centrifugal force at the center of the Earth, C, is equal to that of the gravity due to the Moon. Both forces point in opposite direction, so that they balance each other and cancel. Otherwise, the Earth and the Moon cannot stay resting in the co-rotating frame. The red arrows are the vector sum of the gravity due to the Moon and the centrifugal force, and they depict the resultant tidal force. The magnitude of both red arrows is the same.

The problem is that the centrifugal forces at N, F and C are all assumed to be the same. Without this, the correct tidal force is not obtained. This is the most confusing point to be discussed hereafter. Although we describe that the black arrows in Fig. 3 depict the centrifugal force due to the rotation of the system, it is not correct in the exact sense. Fig. 4 explains the problem, in which G is the common center of gravity and resides inside the Earth.

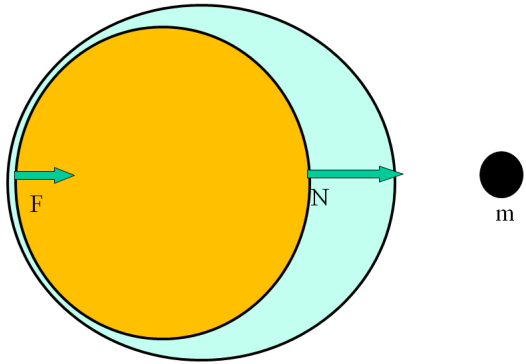

Fig. 1. A wrong picture of the sea' surface on the Earth due to the tidal force. In the figure the brown circle depicts the Earth, the blue ellipsoid the seawater, the black circle the Moon, the green arrows the gravity due to the Moon; N is the near side and F the far side relative to the Moon, respectively.

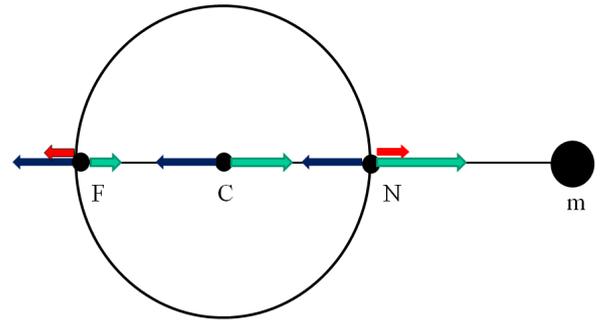

Fig. 3. The correct picture of an explanation of the tidal force. The black arrows show a constant centrifugal force due to the co-rotation of the Earth-Moon system. The red arrows show the tidal force, which are the residuals of the Moon's gravity and the centrifugal force. The magnitudes of the tidal force at N and F are the same, but they point in opposite directions.

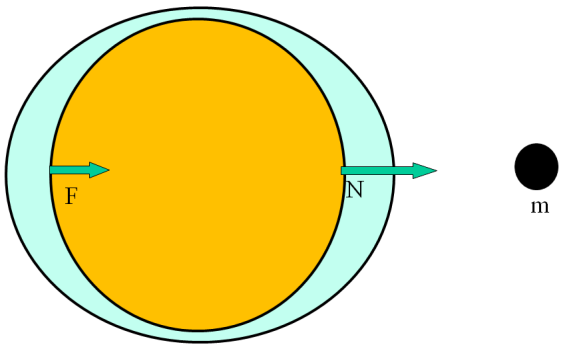

Fig. 2. The correct picture of the equipotential surface or the sea surface

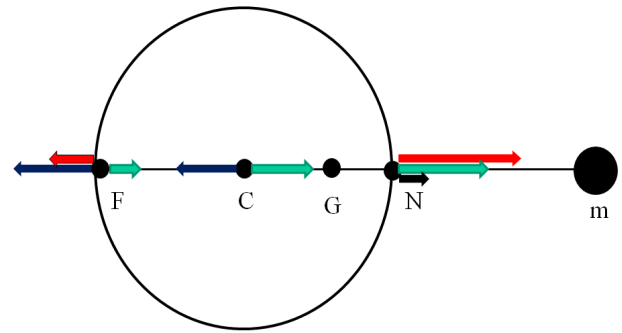

Fig.4. The wrong picture of the explanation of the tidal force, although the centrifugal forces are correctly displayed. In the figure, G is the common center of gravity of the system of the Earth and the Moon, which resides in the Earth. The correct centrifugal forces are depicted by black arrows. At N, the centrifugal force points towards the Moon. These centrifugal forces contradict with those shown in Fig. 3.

Consider any point in this rotating frame. The magnitude of the centrifugal force at this point is proportional to the distance between the point and G, and the centrifugal force points away from G. The correct centrifugal force should be, therefore, the black arrows shown in Fig. 4 rather than those in Fig. 3. However, if this were the case, the correct tidal force cannot be obtained. This is the dilemma to be clarified hereafter.

Fig. 5 explains the correct picture. This figure shows the orbital plane of the Earth-Moon system. The points C, m, G and A are the center of the Earth, the Moon, the common center of gravity and an arbitrary point on the surface of the Earth, as before. Note that the point A is not fixed to the Earth, but on the co-rotating frame. The magnitude of the centrifugal force due to the rotation of the system is proportional to the distance GA and directs from G to A as is shown by the arrow GA.

The vector $\vec{GA}$ can be decomposed as follows:

$$\vec{GA} = \vec{GC} + \vec{CA} \qquad (2.1)$$

The vector $\vec{GC}$ is always the same for any point A. The vector $\vec{CA}$ is axi-symmetric about the center of the Earth C and directs outward. This part does not contribute to the tidal force, and can be combined with the Earth's gravity. Thus, only the constant vector $\vec{GC}$ contributes to the tidal force. Therefore, the explanation in Fig. 3 is correct.

3. The case of the Earth/Moon system without rotation

So far, we have given a qualitative discussion. Hereafter, we give a more quantitative version, which will be divided into three parts. In the first part, we consider the case in which the Earth and the Moon do not revolve each other. They are either in a free-fall state or supported by some mean, e.g., a rod. The one-dimensional treatment is sufficient in this case. In the second part, the Earth and the Moon revolve about the common center of gravity. We restrict ourselves in the two-dimensional rotational plane in this part. The full three-dimensional treatment will be given in the third part.

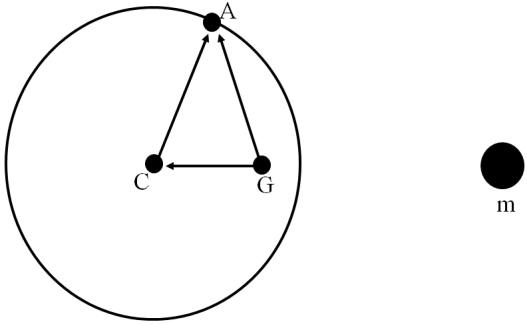

Fig.5. The rotational plane involving the center of the Earth C, the common center of gravity G and the Moon m. A is an arbitrary point on the rotational plane. The centrifugal force vector $\vec{GA}$ can be decomposed in the common part $\vec{GC}$ and the radial part $\vec{CA}$, the last of which does not contribute to the tidal force because of its axi-symmetric nature.

In Fig. 3 the black circle m on the right is the secondary object with mass $m$. The circle on the left is an extended object such as the Earth, the center of which is C. Draw a line connecting C and m. Consider two points on the surface of the Earth. The point N is near and the point F is far with respect to the Moon. Let us denote the distance between C and m by $R$, and the radius of the Earth by $r$. The gravitational acceleration at N, C and F exerted by the Moon are described as follows:

$$\left.\begin{array}{l} g_N = \dfrac{Gm}{(R-r)^2} \\[4pt] g_C = \dfrac{Gm}{R^2} \\[4pt] g_F = \dfrac{Gm}{(R+r)^2} \end{array}\right\} \quad (3.1)$$

The green arrows in Fig. 3 represent these gravitational accelerations.

The tidal acceleration $g_{NC}$ at N is the difference of the gravitational accelerations at N and C

$$g_{NC} = g_N - g_C = \frac{Gm}{(R-r)^2} - \frac{Gm}{R^2}$$
$$= \frac{Gm}{R^2}\left[\left(1-\frac{r}{R}\right)^{-2} - 1\right] = \frac{2Gm}{R^3}r + O\!\left(\frac{r^2}{R^2}\right) \quad (3.2)$$

The first term on the RHS is the tidal acceleration at N, and the rest is the higher order term.

Likewise the tidal acceleration at F is given by:

$$g_{FC} = -\frac{2Gm}{R^3}r + O\!\left(\frac{r^2}{R^2}\right) \quad (3.3)$$

The magnitude of the RHS of (3.3) is the same as that of (3.2), but the sign is opposite as is expected.

The tidal acceleration can be obtained, as well, by differentiating the gravity due to the Moon at C:

$$\Delta g_C = \frac{dg_C}{dR}\Delta R = -\frac{2Gm}{R^3}\Delta R \quad (3.4)$$

If the Earth and the Moon are falling freely in the inertial frame, an observer at C sees the tidal force shown by the red arrows in Fig. 3. In this case, there is no centrifugal force, and the black arrows are interpreted as the inertial forces caused by the accelerating observer at C.

If the Earth and the Moon are supported by, say a rod, so that they do not fall freely, the resistance force, $-g_C$, acts uniformly on the Earth. The tidal acceleration (3.2) and (3.3) are the same in this case as well.

### 4. Two-dimensional analysis of the revolving Earth/Moon system

We consider the case in which both the Earth and the Moon are revolving about their common center of gravity G, the angular velocity of which is $\Omega$. The common center of gravity resides inside the Earth as is shown in Fig. 4, but this fact is not essential to the following discussion. In this case we have to include the centrifugal force acting on a test particle.

Consider a point A on the surface of the Earth. Note that the point A stays still on the co-rotating frame and is not fixed to the Earth's surface. We have to take into account the centrifugal force due to the rotation of the frame; the Coriolis force does not operate.

In the previous section, we restricted ourselves to the line joining C and m. In the following, we consider the two-dimensional rotational plane. As shown in Fig. 6, we denote the distance between C and A as $r$, the distance between C and m as $R$, and the distance between A and m as $r_1$. The angle between CA and Cm is $\psi$.

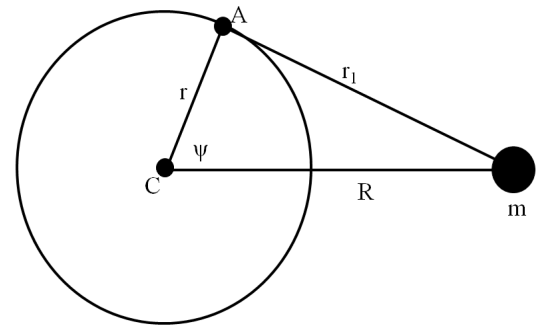

Fig. 6.  Geometry of calculation of the tidal potential.

We have the following equation:

$$r_1^2 = R^2 - 2rR\cos\psi + r^2 \quad (4.1)$$

We calculate $1/r_1$ in order to calculate the tidal potential as

$$r_1^{-1} = \frac{1}{R}\left[1 - 2\frac{r}{R}\cos\psi + \left(\frac{r}{R}\right)^2\right]^{-1/2} \quad (4.2)$$

Assuming $r$ is sufficiently smaller than $R$, we have

$$r_1^{-1} = \frac{1}{R}\left[1 + \frac{r}{R}\cos\psi + \frac{1}{2}\left(\frac{r}{R}\right)^2(3\cos^2\psi - 1)\right] \quad (4.3)$$

The tidal potential at the point A is

$$V_m = -\frac{Gm}{r_1} = -\frac{Gm}{R}\left[1 + \frac{r}{R}\cos\psi + \frac{1}{2}\left(\frac{r}{R}\right)^2(3\cos^2\psi - 1)\right] \quad (4.4)$$

The force is the gradient of the potential. Since the first term in the square bracket in (4.4) is constant, it does not contribute to the tidal force and we may ignore it.

We may set the *x*-axis parallel to Cm, then we have

$$x = r\cos\psi \quad (4.5)$$

The force arising from the second term in the square bracket is

$$g_x = -\frac{\partial V}{\partial x} = \frac{Gm}{R^2} \quad (4.6)$$

This is constant and corresponds to the gravity due to the moon at C.

Finally the tidal potential $V_t$ is the third term in the square bracket in (4.4):

$$V_t = -\frac{Gm}{2R^3}r^2(3\cos^2\psi - 1) \quad (4.7)$$

By calculating the gradient of the tidal potential, we have the tidal acceleration, which can be decomposed into the horizontal component $g_H$ on the Earth's surface and the radial component $g_r$ perpendicular to the Earth's surface.

$$\begin{aligned} g_r &= -\frac{\partial V_t}{\partial r} = \frac{Gm}{R^3}r(3\cos^2\psi - 1) \\ g_H &= -\frac{1}{r}\frac{\partial V_t}{\partial \psi} = -\frac{3Gm}{2R^3}r\sin 2\psi \end{aligned} \quad (4.8)$$

Now we consider the centrifugal force. Let a point A $(x, y)$ be on the rotational plane and on the surface of the Earth as well. The rotational axis of the system is perpendicular to the rotational plane, and it passes through G. Here, the coordinates of G are $(x_G, 0)$.

The centrifugal force at A can be calculated as follows. It directs from G to A, and its magnitude is

$$g_C = r_2\Omega^2 \quad (4.9)$$

Here $r_2$ is the length of GA, which is calculated by the following equation:

$$r_2^2 = (x - x_G)^2 + y^2 \quad (4.10)$$

Therefore, the centrifugal potential $V_C$ is

$$V_C = -\frac{\Omega^2}{2}\left[(x - x_G)^2 + y^2\right] \quad (4.11)$$

The decomposition of the centrifugal potential discussed before can be understood in the following manner:

$$\begin{aligned} V_C &= -\frac{\Omega^2}{2}\left[x^2 + y^2 - 2xx_G + x_G^2\right] \\ &= V_1 + V_2 - \frac{\Omega^2}{2}x_G^2 \end{aligned} \quad (4.12)$$

$$V_1 = -\frac{\Omega^2}{2}(x^2 + y^2) \quad (4.13)$$

$$V_2 = \Omega^2 x_G x \quad (4.14)$$

The last term on the RHS of (4.12) is a constant and does not contribute to the force, and it can be ignored. The term $V_1$ is the centrifugal potential originated from the rotation of the Earth, with angular speed $\Omega$. This term does not contribute to the tidal force either.

Differentiating $V_2$ with respect to $x$, we have a constant acceleration GC. This component of the centrifugal force is constant which is what we want to prove in the present paper:

$$g_2 = -\frac{\partial V_2}{\partial x} = -\Omega^2 x_G \quad (4.15)$$

This force balances with the gravitational acceleration due to the Moon, i.e. (4.6), and we obtain

$$\frac{Gm}{R^2} - \Omega^2 x_G = 0 \quad (4.16)$$

To express the centrifugal potential as the sum of $V_1$ and $V_2$ explains the decomposition of the centrifugal acceleration explained in Fig. 5.

### 5. Three-dimensional analysis

As can be seen in Fig. 6, the tidal potential is ax-symmetric about the *x*-axis. We introduce the three-dimensional Cartesian coordinate $(x, y, z)$ and the polar coordinates $(r, \theta, \phi)$. The origin is at C. The positive direction of the *x*-axis is rightward in the figure. The *y*-axis is in the rotational plane. The *z*-axis is perpendicular to the rotational plane and passes through C. The relation between the Cartesian coordinate and the polar coordinates is

$$\begin{aligned} x &= r\sin\theta\cos\varphi \\ y &= r\sin\theta\sin\varphi \\ z &= r\cos\theta \end{aligned} \quad (5.1)$$

Comparing this with (4.5), we have

$$\cos\psi = \sin\theta\cos\varphi \quad (5.2)$$

The equation (4.7) can be written as

$$V_t = -\frac{Gm}{2R^3}r^2(3\sin^2\theta\cos^2\varphi - 1) \quad (5.3)$$

The centrifugal potential $V_1$ can be expressed in terms of the polar coordinate as:

$$V_1 = -\frac{1}{2}\Omega^2 r^2 \sin^2\theta \qquad (5.12)$$

After all, the total potential is the sum of that due to Earth, the tidal potential and the centrifugal one;

$$V = -\frac{GM}{r} - \frac{Gm}{2R^3}r^2\left(3\sin^2\theta\cos^2\varphi - 1\right) - \frac{1}{2}\Omega^2 r^2 \sin^2\theta \quad (5.13)$$

This form of the potential is nothing but an approximate form of the more general Roche potential.

References
(1) http://en.wikipedia.org/wiki/Tidal_force
(2) http://en.wikipedia.org/?title=Talk:Tidal_force